\documentclass{emulateapj}
\usepackage{natbib}
\def\teff{T$_{\rm eff}$}
\def\logg{log(g)}

\def\kms{km\,s$^{-1}$}

\begin{document}
\title{Light-element abundance variations at low metallicity: the 
globular cluster NGC 5466
\footnote{Based in part on data obtained at the W.M. Keck Observatory, which is operated as a scientific partnership among the California Institute of Technology, the University of California and the National Aeronautics and Space Administration. The Observatory was made possible by the generous financial support of the W.M. Keck Foundation. }}

\bigskip
\author{Matthew Shetrone} 
\affil{McDonald Observatory, University of Texas at Austin, HC75 Box 1337-MCD, Fort Davis, TX 79734, USA}
\email{shetrone@astro.as.utexas.edu}

\author{Sarah L. Martell} 
\affil{Astronomisches Rechen-Institut, Zentrum f\"{u}r Astronomie der Universit\"{a}t Heidelberg, M\"{o}nchhofstr. 12-14, D-69120 Heidelberg, Germany}
\email{martell@ari.uni-heidelberg.de}

\author{Rachel Wilkerson\footnote{REU student at McDonald Observatory}} 
\affil{Center for Astrophysics, Space Physics and Engineering Research
One Baylor Place \#97310, Baylor University Waco, TX 76798-7310, USA}
\email{rachelw52@gmail.com}

\author{Joshua Adams} 
\affil{Astronomy Department, University of Texas at Austin, Austin, TX 78712, USA}
\email{jjadams@astro.as.utexas.edu}

\author{Michael H. Siegel} 
\affil{Department of Astronomy and Astrophysics, Pennsylvania State University, 525 Davey Laboratory, State College, PA 16801, USA}
\email{siegel@astro.psu.edu}

\author{Graeme H. Smith} 
\affil{University of California Observatories / Lick Observatory, Department of Astronomy \& Astrophysics, UC Santa Cruz, 1156 High St., Santa Cruz, CA 95064, USA}
\email{graeme@ucolick.org}

\and

\author{Howard E. Bond} 
\affil{Space Telescope Science Institute, 3700 San Martin Dr, Baltimore, MD 21218, USA}
\email{bond@stsci.edu}

\bigskip
\begin{abstract} 
\noindent
We present low-resolution (R$\simeq 850$) spectra for 67 asymptotic giant 
branch, horizontal branch and
red giant branch (RGB) stars in the low-metallicity globular cluster NGC 5466, taken with
the VIRUS-P integral-field spectrograph at the 2.7-m Harlan J. Smith
telescope at McDonald Observatory. Sixty-six stars are confirmed, and one
rejected, as cluster members based on radial
velocity, which we measure to an accuracy of 16 \kms via
template-matching techniques. CN and CH band strengths have been measured
for 29 RGB and AGB stars in NGC 5466, and the
band strength indices measured from VIRUS-P data show close agreement
with those measured from Keck/LRIS spectra previously taken of five of
our target stars. We also determine carbon abundances from comparisons with
synthetic spectra. The RGB stars in our data set cover a range in absolute $V$ magnitude from $+2$ to $-3$, which permits us to study the rate of carbon depletion on the giant branch as well as the point of its onset. The data show a clear decline in carbon abundance with
rising luminosity above the luminosity function ``bump''  on the giant
branch, and also a subdued range in CN band strength, suggesting ongoing
internal mixing in individual stars but minor or no primordial
star-to-star variation in light-element abundances.
\end{abstract}


\section{Introduction}
     In all Galactic globular clusters with moderate metallicity 
([Fe/H] $\simeq -1.5$), there are star-to-star variations 
in the strength of the $3883 \hbox{\AA}$ CN absorption band at any given 
evolutionary phase. These CN band strengths anticorrelate with CH band strengths, so that stars with weak CH bands (and therefore relatively low carbon 
abundances) have strong CN bands (and therefore relatively high nitrogen 
abundances). Band-strength variations were first observed in spectra of red 
giants in globular clusters (e.g., Norris \& Freeman 1979\nocite{NF79}, Suntzeff 1981\nocite{S81}), 
but are also found in subgiants and on the main sequence (e.g., 
Briley et al. 2004\nocite{BCS04}), indicating that they are a result of primordial 
enrichment rather than an evolutionary effect.

     Variations in the broad CN and CH features are just the most readily 
observed part of a larger light-element abundance pattern, a division of 
globular cluster stars into one group with typical Population II abundances 
and another that is relatively enhanced in nitrogen, sodium, and aluminum, 
and depleted in carbon, oxygen, and magnesium. These abundance divisions 
are usually studied in the anticorrelated pairs of carbon and nitrogen or 
oxygen and sodium (see, e.g., Kraft 1994\nocite{K94}; Gratton et al. 2004\nocite{GSC04}; 
Carretta et al. 2009\nocite{C09}). The magnesium-aluminum anticorrelation is more murky: 
although some researchers find Mg-Al anticorrelations in individual globular 
clusters (e.g., Shetrone 1996\nocite{S96}; Gratton et al. 2001\nocite{GBB01}), the abundance 
distributions are not as clearly bimodal as in the case of nitrogen, and 
the origin of Mg-poor, Al-rich stars is more difficult to fit into the 
general framework of enrichment by moderate-mass AGB stars (e.g., 
D'Ercole et al. 2008\nocite{DVD08}).

     This paper presents a study of the behavior of CN and CH band strengths 
and [C/Fe] abundances in red giant stars in the low-metallicity globular 
cluster NGC 5466, which has [Fe/H] $\simeq 2.2$ (Harris 1996, 
2003 revision\nocite{H96}). Studies based on CN and CH band strength are more difficult 
in low-metallicity clusters, where the CN variation visible in the spectra 
can be quite small despite variations in [N/Fe] abundance as large as in 
higher-metallicity globular clusters (e.g., M53, Martell et al. 2008a\nocite{MSB08a}; 
M55, Briley et al. 1993\nocite{BSHB93}). The globular clusters with metallicities 
similar to NGC 5466 in which light-element abundance inhomogeneities have 
been most extensively studied are M92 and M15. The red giants in both 
objects exhibit a progressive decline in [C/Fe] with advancing evolution 
above the magnitude level of the horizontal branch (HB; Carbon et al. 1982\nocite{Carbon1982}; 
Trefzger et al. 1983\nocite{T1983}; Langer et al. 1986\nocite{Langer1986}; Bellman et al. 2001\nocite{Bellman2001}). Giants 
in both clusters typically exhibit weak $\lambda$3883 CN bands 
(Carbon et al. 1982\nocite{Carbon1982}; Trefzger et al. 1983\nocite{T1983}), and do not exhibit bimodal 
CN distributions like those in more metal-rich clusters although a handful 
of stars with enhanced CN have been discovered in M15 
(Langer, Suntzeff, \& Kraft 1992\nocite{Langer1992}; Lee 2000\nocite{L00}). Despite their lack of 
CN-strong giants, both M92 and M15 exhibit anticorrelated O-N, O-Na 
or O-Al variations, or correlated N-Na, of the type that are commonplace 
in globular clusters (Norris \& Pilachowski 1985\nocite{Norris1985}; Sneden et al. 1991\nocite{Sneden1991}, 1997\nocite{Sneden1997}; 
Shetrone 1996\nocite{S96}). In addition, Cohen, Briley \& Stetson (2005)\nocite{Cohen2005} discovered 
that star-to-star differences in [C/Fe] and [N/Fe] abundances exist among 
stars near the base of the red giant branch (RGB) in M15, and the C and N 
abundances tend to be anticorrelated. These efforts have shown that for 
very metal-poor stars the use of CN to determine which stars carry the 
anticorrelated light-element abundance pattern can be problematic, and 
that anticorrelated variations in light-element abundances do exist in 
low-metallicity globular clusters despite their lack of obvious CN 
band strength variation.

      If NGC 5466 does contain the same primordial variations in 
light-element abundances as are found in higher-metallicity globular 
clusters, the resulting range in CN and CH band strength will be muted by 
the low overall metallicity of the cluster, perhaps to a degree where it 
is difficult to distinguish a bimodal but closely spaced CN band strength 
distribution from a broad but unimodal distribution. If NGC 5466 does not 
contain primordial abundance variations, that would imply that primordial 
enrichment is not a universal process in low-metallicity globular clusters 
the way it is in higher-metallicity clusters. A metallicity limit on 
primordial light-element enrichment offers some insight into the primordial 
enrichment process, and into the larger and still-developing picture of 
chemical complexity within individual globular clusters (e.g., Piotto 2008\nocite{P08}).

\section{Observations and Reduction}

We determined photometry for members of NGC 5466 based on CCD frames 
obtained by HEB with the Kitt Peak National Observatory 0.9 m telescope 
on 1997 May 9.  The T2KA chip at the Ritchey-Chretien focus provides 
a $23'\times23'$ field, which encloses most of the cluster, which has a 
tidal radius of $34'$, according to Harris (1996)\nocite{H96}.  We used $uBVI$ filters, as defined by Bond (2005)\nocite{Bond05}, with
exposure times of 600, 45, 45, and 60~s, respectively, under photometric
conditions. We calibrated the photometry to the network of $uBVI$ standard stars
established by Siegel \& Bond (2005)\nocite{SB05}.  
Data were reduced using the IRAF CCDPROC pipeline and photometry was measured 
with DAOPHOT/ALLSTAR (Stetson 1987, 1994\nocite{Stetson87}\nocite{Stetson94}). 
DAOGROW (Stetson 1990\nocite{Stetson90}) was used to perform curve-of-growth 
fitting for
aperture correction on both program and standard stars. The raw photometry 
was calibrated using the iterative matrix inversion technique described in 
Siegel et al. (2002)\nocite{S02} to translate the photometry to the 
standard system of Landolt (1992)\nocite{Landolt92} 
and Siegel \& Bond (2005)\nocite{SB05}.  
Table \ref{t1} lists identification numbers, positions, and 
photometry for the stars observed with VIRUS-P.  
The star identification numbers and coordinates are from an
unpublished photometric catalog developed by M.H.S. from the analysis described
above.  The typical errors on the B, V, and I photometry are 0.030, 0.022 
and 0.032 mag, respectively.

\begin{figure}
\resizebox{\hsize}{!}{\includegraphics{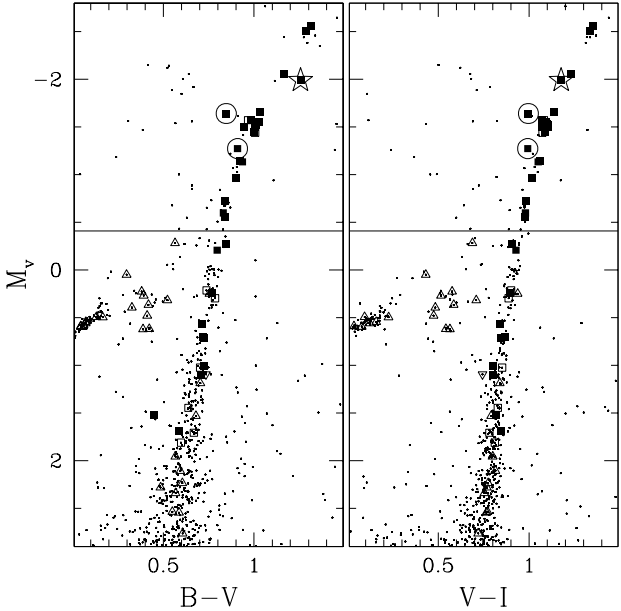}}
\caption{
Color magnitude diagram for the
photometric sample of NGC 5466. Filled squares represent the
spectroscopic sample, and open squares show stars with high enough
S/N for spectroscopic analysis but the possibility of spectroscopic
contamination from a nearby star. The open triangles denote
HB stars and red giants with insufficient S/N for spectral
analysis. The single upside down triangle represents star 1980, a
strong-lined RV non-member. Small points represent stars not observed as part of this VIRUS-P campaign. Filled squares outlined by larger symbols represent AGB stars (outlined with circles) and a probable CH star (outlined with a star).
\label{fig1}}
\end{figure}

On 2009 March 3--5 and July 28--29,
we observed four fields in NGC 5466 for a total of 22 hr with the
visible integral-field replicable unit spectrograph prototype (VIRUS-P,
Hill et al. 2008\nocite{Hill08}) on the 2.7m Harlan J. Smith Telescope at 
McDonald Observatory.  
VIRUS-P is a spectrograph with an integral field 
unit (IFU) of 246 fibers in a hexagonal, one third fill factor, close 
pack pattern. The fibers have a projected diameter of $4.1\arcsec$ and 
have centers astrometrically calibrated to $0.7\arcsec$ relative to a 
fixed, offset guiding camera via open cluster observations.   The field 
of view for the IFU is 2.89 square arcmin and with a set of three 7.14"
dithers the filling factor is 100\%.  
The VP1 grating was used, giving an instrumental FWHM between 5 and $6\hbox{\AA}$ 
over the wavelength range of $3500-5800\hbox{\AA}$. Arc lamp calibration 
produced a wavelength solution with an rms accuracy of $0.05\hbox{\AA}$. 
Our procedure during observing was to dither the telescope and take three
1200 s exposures per field to fill in the spaces between the IFU fibers.   
Three of the fields had four sets of dithers, but the July observing run was 
less successful due to weather and only a single dither set was taken 
for the final field. 

A custom software pipeline developed for the instrument 
(Adams 2010, in preparation)
was used in all reductions. The notable features are that no interpolation 
was performed on the data to avoid correlated noise and that background 
subtraction was done between differently sampled fibers by fitting 
B-spline models (Dierckx 1993\nocite{Dier93}) in a running boxcar of 31 chip-adjacent 
fibers in a manner similar to optimal longslit sky subtraction methods 
(Kelson 2003\nocite{Kels03}). The NGC 5466 pointing was sparse enough to allow 
self-sky-subtraction.   The spectra were flux calibrated by observing
a single flux standard on each night using the same dithering pattern 
used for the science targets.  
The flux solution was then applied to every fiber in every exposure 
for that night.   Based on many tens of flux standards taken during 
other VIRUS-P projects the absolute flux calibration precision is
8.5\% and wavelength independent for full dither sets.   The data reduction
code does not correct for differential atmospheric refraction (DAR) because
there is very little impact from DAR with these
large fibers when they are summed over full dither sets.
Fiber spectra were then combined using the IRAF task
{\it scombine} to form a single spectrum for each star.  For some stars, this
was done by combining a single fiber from all visits (usually four) 
to a specific dither position; for
other stars, those falling near the edge of a fiber, it may have
required combining different fibers on different 
dithers (perhaps 8 or 12 spectra).

\begin{figure}
\resizebox{\hsize}{!}{\includegraphics{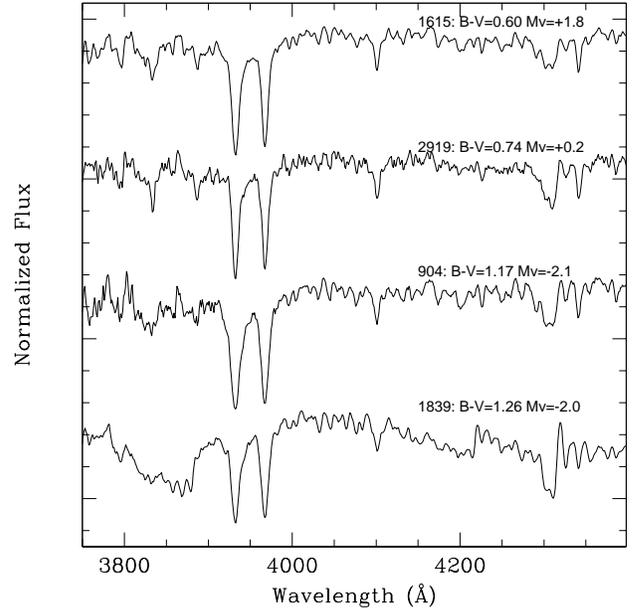}}
\caption{
Four spectra from our spectroscopic sample as examples of the characteristics of VIRUS-P spectra. The top three spectra
were chosen to span the range of absolute magnitude in our data
set. The lowest spectrum, that of star 1839, has very strong CN and
CH bands. 1839 also sits to the red of the RGB in the ($V,B-V$) CMD, and is
probably a CH star. 
\label{fig2}}
\end{figure}

Figure \ref{fig1} is a color-magnitude diagram for NGC 5466, with different
symbols denoting various parts of our data set: stars not observed in
this particular VIRUS-P pointing are small points, and the spectroscopic
sample (unblended asymptotic giant branch (AGB) 
and RGB stars with high enough signal-to-noise 
for index measurements) are shown as filled squares. Our standard for a
spectroscopically unblended star is that there are no other stars within
5 arcsec down to 3 mag below the sample star's B
magnitude. Open squares represent stars with sufficient signal-to-noise (S/N) 
for
spectroscopic analysis but possible spectroscopic blends, and open
triangles represent HB stars and RGB stars with low S/N. 
One strong-lined radial velocity nonmember (star 1980) is shown as an
inverted triangle. The two filled squares outlined with large circles 
are stars 1398 and 2483, and are likely to be AGB stars based on their 
positions to the blue of the giant branch. The filled square outlined 
with a large star is star 1839, which has strong CN and CH features 
and is likely a CH star. Although 1839 lies on the giant branch in the 
($V, V-I$) color-magnitude diagram, the strong absorption in the UV, due
to extra CN opacity and the Bond-Neff effect (Bond \& Neff 1969\nocite{BN69}), 
makes its ($B-V$) color distinctly red relative to the giant branch. 
The distance modulus and reddening for NGC 5466 are ($m-M$)$_{V}=16.15$
and E($B-V$)=0.023 (Dotter et al. 2010\nocite{D10}).

Four typical spectra are shown in Figure \ref{fig2}, to provide a sense 
for the quality and characteristics of the data. The top three spectra are 
a series in luminosity spanning the range of our data set, with the
faintest at the top. The lowest spectrum is of star 1839, and shows the 
unusually strong CN and CH bands. 

\subsection{The LRIS Comparison Sample}

Spectra were acquired for six RGB stars in NGC 5466 by GHS in 2003, using 
the LRIS double spectrograph (Oke et al. 1995\nocite{Oke95}) 
at the W.M.~Keck Observatory 
on Mauna Kea. All light was directed to the blue arm of the spectrograph 
by the use of a mirror in place of a dichroic. The 400/3400 grism was 
used, resulting in a typical resolution of $7\hbox{\AA}$ and a pixel 
spacing of $1\hbox{\AA}$.  Exposure times were  between 420 and 600 s 
producing a mean S/N of 225 pixel$^{-1}$ in the wavelength 
range $4000\hbox{\AA} \le \lambda \le 4100\hbox{\AA}$. All reductions 
from flat field division through flux calibration were carried out using 
the XIDL routines developed by J.X. Prochaska.

\begin{figure}
\resizebox{\hsize}{!}{\includegraphics[angle=-90]{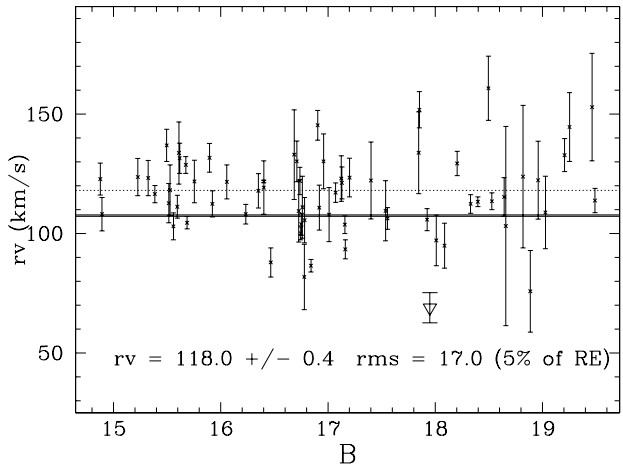}}
\caption{
Radial velocities for all stars in our photometric sample, with a horizontal dashed line marking the mean velocity of $118.0$ \kms. The horizontal solid line represents the mean cluster radial velocity values taken from Peterson \& Latham (1986) and Pryor et al. (1991). The error on our mean velocity is $0.4$ \kms, and the rms is $17.0$ \kms. This rms is 5\% of a resolution element with VIRUS-P.  The single inverted triangle represents star 1980, a strong-lined star that is also a radial velocity non-member. 
\label{fig3}}
\end{figure}

\section{Radial Velocity and Membership}

For radial velocity determinations, data were cross-correlated with a 
template spectrum: for RGB and AGB stars, the template was 
the Hinkle Arcturus atlas (Hinkle et al. 2000\nocite{HA00}) 
convolved with a Gaussian to create 
an $R=850$ spectrum. The HB star template was generated from a synthetic 
spectrum with \teff$=6000$ K, \logg$=3.00$ dex and [Fe/H]$=-2.0$ dex.  

Figure \ref{fig3} shows the distribution of our velocity measurements. The 
weighted average is $118.0 \pm 0.4$ \kms with an rms of $17.0$ \kms. This 
rms is $5$\% of a resolution element with VIRUS-P. More customized template 
spectra may allow improvements in the rms scatter. The mean of literature 
values for the systemic velocity of NGC 5466 is $108$ \kms (Harris 1996\nocite{H96}), 
and solid horizontal lines in Figure 4 show velocities from 
Pryor et al. (1991)\nocite{P91} and Peterson \& Latham (1986)\nocite{PL86}. 
The individual star radial velocities are listed in Table \ref{t1}.
In the following analysis, spectra are individually shifted to rest
wavelengths according to these radial velocities.

\section{Band Strengths and Bimodality}
CN and CH band strengths are typically quantified with indices that
measure the magnitude difference between the integrated flux within
the feature in question and the integrated flux in one or two nearby
continuum regions, in the sense that more absorption in the feature
produces a larger band-strength index. For CN band strength, we measure the index 
$S(3839)$, defined in Norris et al. (1981)\nocite{N81} as 
\begin{eqnarray}
S(3839) = -2.5 \log \frac{\int_{3883}^{3916}I_{\lambda}
  d\lambda}{\int_{3846}^{3883}I_{\lambda} d\lambda}
\end{eqnarray}
The index we use for CH band strength is $S_{2}(CH)$, defined in
Martell et al. (2008b)\nocite{MSB08b} as
\begin{eqnarray}
S_{2}(CH) = -2.5 \log \frac{\int_{4297}^{4317}I_{\lambda}
  d\lambda}{\int_{4212}^{4242}I_{\lambda} d\lambda +
  \int_{4330}^{4375}I_{\lambda} d\lambda}
\end{eqnarray}

\begin{figure}
\resizebox{\hsize}{!}{\includegraphics[angle=-90]{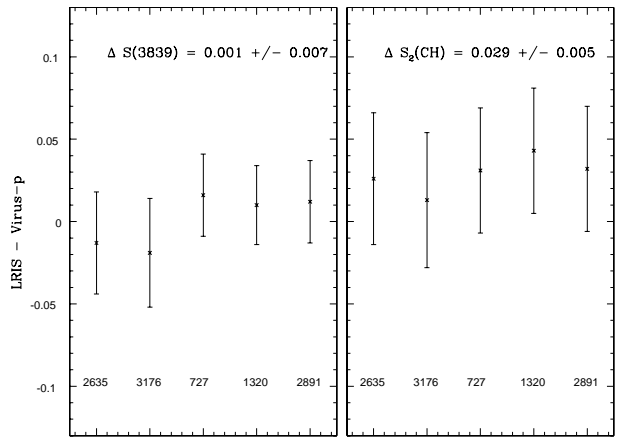}}
\caption{
Comparison of band indices measured from the Keck LRIS spectra
discussed in Section 2.1 and those measured from VIRUS-P spectra, for
the five stars in common to both data sets. 
Identification numbers are given at the bottom of each panel.
\label{fig4}}
\end{figure}

Band-strength indices tend to be designed to be effective for a single stellar population, for example $CH(4300)$ (Harbeck et al. 2003\nocite{HSG03a}, designed for main-sequence stars in 47 Tucanae) and $CH(G)$ (Lee 1999\nocite{L99}, designed for bright red giants in M3), and can be difficult to apply outside the parameter space for which they were intended. $S_{2}(CH)$ was specifically designed to be sensitive to variations in carbon abundance in bright red giants ($M_{V} \le -1.0$) across a wide range of metallicity, and therefore ought to be valid for the sample at hand.

The errors on our band strengths have two components, errors related to 
the fluxing and errors related to the S/N.  
S/N errors were calculated by propagating the Poisson noise 
errors through the formulas for the band indices.  To estimate the 
errors in our fluxing we tried several techniques to reflux the data
including using synthetic spectra matched to each star's photometric 
temperature and using average flux distributions for color bins 
along the giant branch.  The two sources of error were added in 
quadrature, and the total errors 
are reported in Table \ref{t2} as $\sigma_{S}$ and $\sigma_{S2}$ 
for $S(3839)$ and $S_{2}(CH)$, respectively.   For the March observing run
the S/N in the $S_{2}(CH)$ band was quite high and the fluxing errors 
dominate while for the July observing run, where only a single dither set 
was taken for the final fourth field, the Poisson noise contribution 
plays a significant contribution to the total error.   For $S(3839)$,
the Poisson error tends to dominate the total error in all cases.
Errors in radial velocity are typically 17 \kms, or 5\% of a pixel, as
mentioned in Section 3, and do not have a significant effect on measured
band strengths. Shifting our wavelengths to the blue or red by 17 \kms
changes the measured $S_{2}(CH)$ by 0.001 mag, which is small
relative to the fluxing- and noise-related errors in $S_{2}(CH)$.

\begin{figure}
\resizebox{\hsize}{!}{\includegraphics{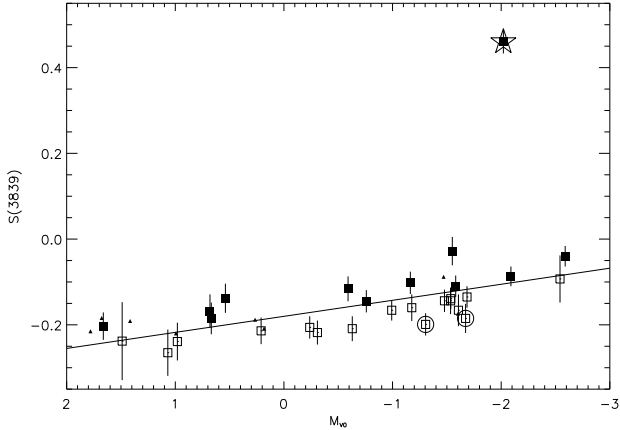}}
\caption{
CN band-strength index $S(3839)$ vs. absolute $V$ magnitude, for
all stars in our spectroscopic sample. The solid line is a
least-squares fit, and we classify stars that lie above the line
(shown as filled squares) as relatively CN-strong and stars that lie below the
line (drawn as open squares) as relatively CN-weak. Small triangles
denote stars with sufficient S/N for spectroscopic analysis but
possible spectroscopic blending. The clear outlier, outlined
with a large open star, is star
1839, noted in Figure 2 as a probable CH star. The two points
outlined with large open circles (stars 1398 and 2483) lie to the blue of the RGB in Figure
1, and are likely AGB stars.
\label{fig5}}
\end{figure}

Five of the stars in the LRIS data set are also in the VIRUS-P spectroscopic sample, and the CN and CH band strengths measured from the LRIS spectra are quite similar to band strengths
measured from the VIRUS-P data. Figure \ref{fig4} shows $\Delta S(3839)$ and
$\Delta S_{2}(CH)$ for the stars in common between the two data sets, in the sense LRIS$-$VIRUS-P, and differences
between the two data sets are quite small.   The band strengths given 
in Table \ref{t2} come exclusively from the VIRUS-P data.

Since the $3883 \hbox{\AA}$ CN band strength can be used as a proxy for
nitrogen abundance, and the $4320 \hbox{\AA}$ CH band strength for
carbon abundance, a comparison of CN and CH band strength in a globular
cluster ought to reveal anticorrelated abundance behavior. As can be
seen in Figure \ref{fig5}, the CN band strength in our sample
rises slightly with rising luminosity, but does not show a large range
at fixed $M_{V}$. The clear outlier, with a large open star drawn
around it, is star 1839, the star with the unusually strong CN and CH
features in Figure 2. It is excluded from the rest of our band strength
and abundance analysis. 

\begin{figure}
\resizebox{\hsize}{!}{\includegraphics{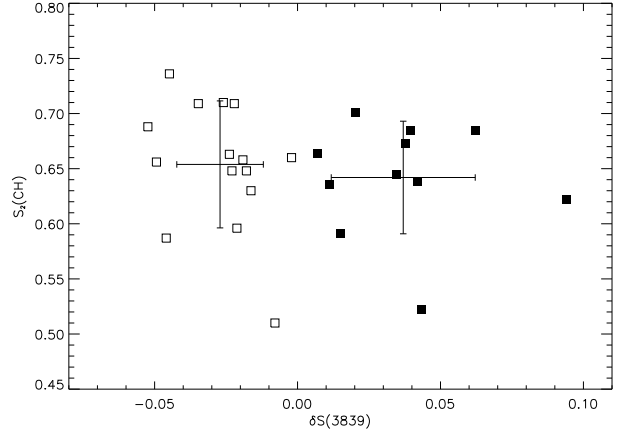}}
\caption{
CN band-strength index $\delta S(3839)$ vs. CH band-strength index
$S_{2}(CH)$, for the spectroscopic sample. The symbols are the same
as in Figure 5. The large crosses are centered at the mean values of
$\delta S(3839)$ and $S_{2}(CH)$ for the CN-weak and CN-strong groups,
and their sizes are the standard deviations of the indices. CN and CH
band strengths are typically anticorrelated in globular cluster stars, indicating
anticorrelated carbon and nitrogen abundances. The lack of strong
anticorrelation here is the result of a smaller than typical range in
CN band strength in NGC 5466, and possibly a smaller range in [C/Fe]
and [N/Fe] than in higher-metallicity globular clusters.
\label{fig6}}
\end{figure}

The $S(3839)$ band strengths are correlated, to first order, with 
luminosity. To better distinguish stars that are CN-weak from stars 
that are CN-strong, we fit a linear relationship between $S(3839)$ 
and $M_{V}$ and evaluated each star based on its vertical distance 
from this linear 
relationship  ($\delta S(3839)$).
We divide the sample into relatively CN-weak (negative $\delta S(3839)$, open squares) and CN-strong (positive $\delta S(3839)$, solid squares) stars. Unlike in studies of higher-metallicity
globular clusters (see Norris et al. 1981\nocite{N81} for a typical example),
there is not a clear gap between the relatively CN-weak and CN-strong
groups. In Figure \ref{fig6} we compare $\delta S(3839)$ and $S_{2}(CH)$ band strengths, with the same
symbols (open/solid squares) representing CN class, and there is not a
strong anticorrelation between CN class and CH band strength. The two
large crosses are centered at the mean values for $\delta S(3839)$ and
$S_{2}(CH)$ for the two groups, with their size set to equal the
standard deviations. They are not obviously offset from each other in
$S_{2}(CH)$. This may be
simply an effect of the extremely limited range in CN and CH
band strengths permitted by the low overall metallicity of NGC 5466.

\begin{figure}
\resizebox{\hsize}{!}{\includegraphics{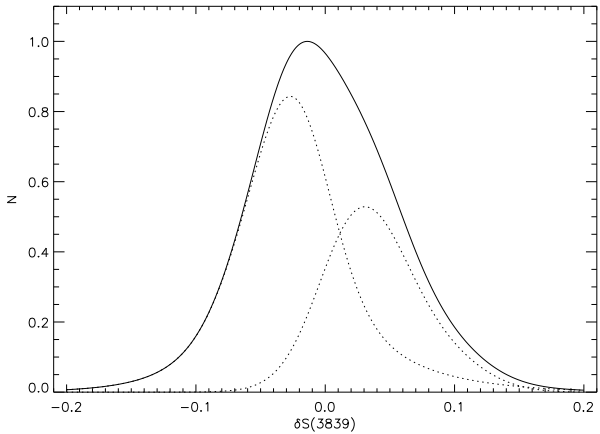}}
\caption{
Solid curve is a generalized histogram for the
luminosity-independent CN band strength measure $\delta S(3839)$
measured for our full spectroscopic sample, and the dashed curves are
generalized histograms for $\delta S(3839)$ in the relatively CN-weak
and CN-strong groups calculated independently. 
\label{fig7}}
\end{figure}

\begin{figure}
\resizebox{\hsize}{!}{\includegraphics{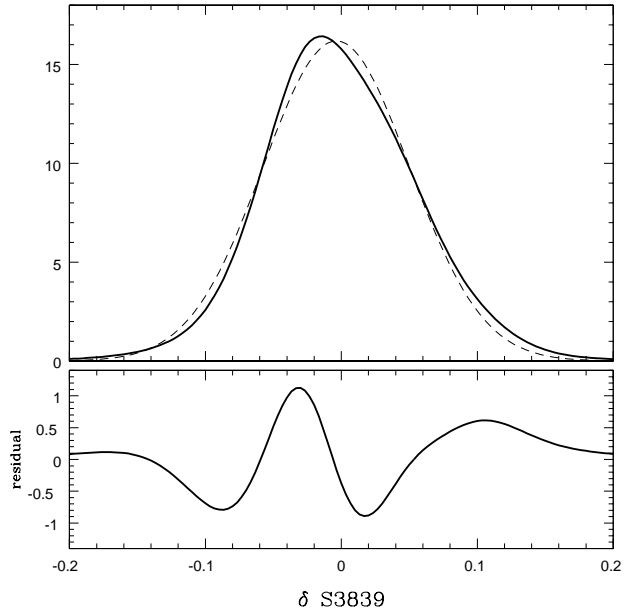}}
\caption{
Upper panel shows the generalized histogram from Figure 7 as a
solid line, with a simple Gaussian fit to the distribution
overplotted as a dotted line. The lower panel shows the difference
between the two curves, and there is a clear two-peaked shape, which
suggests that the data are best represented as two overlapping
Gaussian distributions. 
\label{fig8}}
\end{figure}

Another way to visualize the distribution of CN band strength in our data set is the generalized histogram, which is constructed by representing each star as
a Gaussian in $\delta S(3839)$ with a width equal to the measurement
error $\sigma_{S}$, and then summing the individual Gaussians. 
The result is shown in Figure \ref{fig7}, in which the solid curve 
is the generalized histogram for
the full data set and the dashed curves are the generalized histograms
calculated just for the relatively CN-weak and CN-strong groups. 
The upper panel of Figure \ref{fig8} shows the
single Gaussian that best fits the generalized histogram from Figure
\ref{fig7}, and the lower panel, showing the residual between the best-fit
curve and the data, shows a clearly double-peaked shape.   This 
double-peaked residual suggests the system is better fit by 
two Gaussians.

Based on our analysis of the generalized histogram, 
we claim that NGC 5466 contains two closely spaced
groups in CN band strength, a result of intrinsic ranges in carbon and 
nitrogen abundance and low overall metallicity. The
lack of a gap in CN band strength between the two groups of stars in
Figure \ref{fig5} is seen here as a fairly small distance between the peaks of
the CN-weak and CN-strong generalized histograms. Similarly metal-poor
globular clusters such as M15 (Lee 2000\nocite{L00}) and M53 
(Martell et al. 2008a\nocite{MSB08a}) also exhibit small separations
between their CN-weak and CN-strong groups and a small overall range in $S(3839)$ (among red giants in the three clusters $\delta S(3839)$ has a standard deviation of $0.038$ in NGC 5466, $0.056$ in M53, and $0.037$ in M15). This suggests that the range
of carbon and nitrogen abundance in low-metallicity globular clusters may be smaller than is
typical for more moderate-metallicity clusters. In addition, since the
measurement errors on $S(3839)$ control the widths of the individual
Gaussians comprising the generalized histogram, they can have strong
effects on our interpretation of it, with very large measurement
errors potentially rendering two closely spaced populations
indistinguishable from a single broad distribution. 

To understand the
effects of measurement error on our generalized histogram, we performed a Monte Carlo test exploring how the size of our measurement errors affects the Kolmogorov-Smirnov coefficient $p$, which quantifies the probability that the CN-strong and CN-weak groups were drawn from the same initial population. We rescaled our true measurement errors $\sigma_{S}$ by factors ranging from $0.5$ to $10$, then selected a new value for each data point from within a
Gaussian distribution centered at the measured value of $\delta S(3839)$, with a width of the rescaled measurement
error. This reselection was done for the CN-normal and CN-strong
groups identified in Figure \ref{fig4} independently. The K-S $p$ coefficient 
was then measured as an indicator of whether the two CN band strength groups 
containing reselected points with rescaled errors seemed likely to have been 
drawn from the same initial population. Reselections were done $10^{4}$ times 
for each rescaling of the measurement error, and we find that $p$ is strongly 
concentrated near zero when the rescaling factor is relatively small: for a 
rescaling factor of 1, $99\%$ of realizations have a $p$ less than $0.1$ and 
none have a $p$ larger than $0.61$, and $82\%$ of realizations with a rescaling factor of 3 have a $p$ less than $0.5$. With a rescaling factor of 10, we find that $p$ values are fairly evenly distributed between 0 and 1, with $51\%$ below $p=0.5$. Based on this test, we are fairly confident that our two CN groups really are distinct, even if our errors in $S(3839)$ are underestimated by a factor of 3.

\section{Evolution of Carbon Abundance}
Carbon abundances were determined by spectral synthesis of the G-band
region. For each star, a set of synthetic spectra were created with a
single \teff, \logg, and [Fe/H], and a range of possible values
for [C/Fe], smoothed to the resolution of the VIRUS-P data.  The
synthetic spectra were generated using a modified version of MOOG 2009
(Sneden 1973\nocite{S73}) which generates fluxed spectra instead of normalized spectra.
We used the PLEZ2000 grid of metal-poor model atmospheres (Plez 2000\nocite{P00}).
The temperatures and surface gravities were derived from the photometry
in Table \ref{t1}
using the prescriptions in Ram{\'{\i}}rez \& Mel{\'e}ndez (2005)\nocite{RM05}.  We adopted 
[$\alpha$/Fe] $= 0.3$, $^{12}$C$/^{13}$C$ = 6$ and [Fe/H] $= -2.0$ 
for all stars.
The choice of the carbon isotope ratio does make a small difference 
to the derived total carbon abundance.  At these metallicities, where the 
lines are reasonably weak, a change of $^{12}$C$/^{13}$C from 6 to 50 would
make about 0.06 dex difference to [C/Fe] which is much smaller than the 
quoted errors.  The
synthetic spectrum yielding the residual (in the sense observed - synthetic) 
that is closest to zero in the region of the CH G-band was chosen to have 
the correct carbon abundance.  Table \ref{t2} lists 
ID numbers, VIRUS-P CN and CH band strengths, and carbon
abundances along with their
associated errors for the stars in the 
spectroscopic sample and for those that have good S/N but are possibly 
spectroscopically blended. 

Observational studies of CH band strengths and [C/Fe] abundances in red giants in globular clusters and in the field (e.g., Gratton et al. 2000\nocite{GSCB00}; Smith \& Martell 2003\nocite{SM03}; Smith et al. 2005\nocite{SBH05}) have shown clearly that bright, low-mass red giants in all Galactic environments experience steady carbon depletion as they evolve along the giant branch. This depletion only occurs in stars brighter than the ``bump'' in the RGB luminosity function, and proceeds at a rate that is metallicity-dependent. The homogeneity and wide luminosity range of our spectroscopic sample makes it very useful for careful study of the rate and point of onset of deep mixing in NGC 5466.

\begin{figure}
\resizebox{\hsize}{!}{\includegraphics{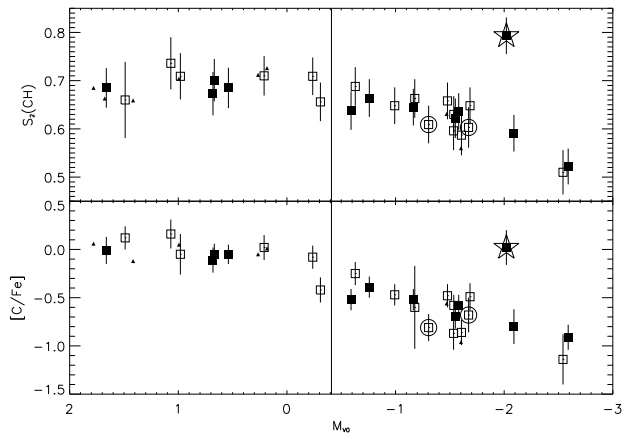}}
\caption{
Upper panel shows the CH band-strength index $S_{2}(CH)$ as a
function of absolute $V$ magnitude, and the usual globular cluster
pattern of a plateau at low luminosity, followed by a decline in
stars brighter than the "bump" in the RGB luminosity function, is
apparent. The vertical line marks the position of the RGB bump at
this metallicity. The lower panel shows our calculated [C/Fe]
abundances, which also exhibit a plateau followed by a steady
evolutionary decline. As discussed in the text, the rate of carbon
depletion measured in NGC 5466 is consistent with the low-metallicity
carbon depletion rate found in Martell et al. (2008c).
\label{fig9}}
\end{figure}

The first dredge-up process that occurs when a star moves on to the RGB (Iben 1965, 1968\nocite{I65}\nocite{I68}) is caused by the base of the convective envelope dropping in to small radius, until hydrogen shell burning ignites, equilibrates, and restores the stability against convection of much of the stellar interior, driving the base of the convective zone to larger radius. This event homogenizes all material in the star outside the minimum radius reached by the convective envelope, and creates a discontinuity in the mean molecular weight $\mu$ at that radius.  This $\mu$ discontinuity is destroyed when the hydrogen-burning shell, constantly moving outward in radius in search of fuel, encounters it. When that happens, the sudden change in the composition of the fuel available to the hydrogen-burning shell causes a brief loop in the star's path through the color-magnitude diagram: it becomes hotter and dimmer before recovering its equilibrium and continuing to evolve along the RGB. This evolutionary stutter causes a 
pile-up of stars in the luminosity function of the RGB, known as the ``RGB bump'' (see, e.g., Fusi Pecci et al. 1990\nocite{FP90}). 
The time at which a red giant's hydrogen-burning shell crosses the $\mu$ discontinuity is a unique function of stellar mass and metallicity, meaning that the absolute $V$ magnitude of the RGB bump can be calculated empirically or theoretically for a given globular cluster. 

This destruction of the $\mu$ barrier permits slow
mixing between the surface and the hydrogen shell, which results in a
continuous decline in surface carbon abundance and an associated
decline in CH band strength starting at the RGB bump and continuing for
as long as a star is on the giant branch (see, e.g., Palacios et al. 2006\nocite{P06}; Denissenkov \& VandenBerg 2003\nocite{DV03}). The structure of the hydrogen-burning shell is also a function of stellar metallicity, and is more compressed in higher-metallicity stars (see, e.g., Sweigart \& Mengel 1979\nocite{SM79} for a discussion of hydrogen-burning shell structure). The more radially extended structure of the hydrogen-burning shell in lower-metallicity stars allows deep mixing currents to carry material more efficiently from the stellar surface into the zone in which the CN cycle is operating, and return carbon-depleted, nitrogen-enhanced material to the surface. The rate of carbon depletion at the surface, then, carries a fair amount of information about the internal structure of the star, and also about the process that drives deep mixing currents (e.g., Eggleton et al. 2008\nocite{EDL08}). 

The upper panel of Figure \ref{fig9} shows CH band strength as a function of
absolute $V$ magnitude, and the expected pattern of a plateau followed by a
steady decline with rising luminosity can be seen. We calculate that
the RGB bump ought to be at $M_{V}=-0.41$, using the relation
$M_{V}^{bump}=0.94$[Fe/H]$+1.5$ found empirically in 
Fusi Pecci et al. 1990\nocite{FP90}, 
and the vertical line in
Figure \ref{fig9} is at that luminosity. It is unclear whether our
data completely agree with this prediction; given the sparseness of
the data in the range $-0.5 \leq M_{V} \leq 0.0$, the corner between
the plateau and the evolutionary decline could be anywhere within that
range. 

The lower panel of Figure \ref{fig9} shows our derived carbon
abundances as a function of absolute $V$ magnitude, and [C/Fe] shows
the same plateau, corner, and decline above the luminosity of the RGB
bump ($M_{V} \simeq -0.4$) as $S_{2}(CH)$ does. The carbon abundance
before the onset of deep mixing is roughly solar, and drops by a dex
by the tip of the giant branch. As was demonstrated in Martell et al. (2008c)\nocite{MSB08c},
carbon depletion in globular cluster red giants proceeds roughly twice
as fast in the low-metallicity clusters M92 and M15 as it does in the
higher-metallicity clusters M5 and NGC 6712. Using Yale-Yonsei
evolutionary tracks (Demarque et al. 2004\nocite{D04}), we determine the age difference
between a star at $M_{V}=-0.41$ (the height of the RGB bump) and
$M_{V}=-1.5$ to be 20 Myr, implying a carbon depletion rate of roughly
$34$ dex Gyr$^{-1}$. This is in good agreement with the result of
Martell et al. (2008c)\nocite{MSB08c}, which found a carbon depletion 
rate of $20-30$ dex Gyr$^{-1}$
for globular clusters with metallicities between $-2.0$ and $-2.5$.

\section{Summary}

We do not see strong star-to-star variations in CN or CH band strength
in NGC 5466, but we do find that a sum of two Gaussians represents
the generalized histogram of CN band strength better than a single
Gaussian curve. This suggests that the underlying $S(3839)$ distribution is
likely to consist of two groups that are distinct but not widely
separated. There is a very weak anticorrelation between CN and CH band
indices, but it is muted even compared to similarly metal-poor
clusters. It may be that NGC 5466 experienced weaker primordial
enrichment than other Galactic globular clusters, and even other
low-metallicity globular clusters. It would be interesting to
investigate whether theoretical scenarios for primordial enrichment of
globular clusters predict varying enrichment efficiencies for
particularly low-metallicity or low-mass clusters.

Although the initial abundances in NGC 5466 may exhibit less of a range 
than abundances in typical globular clusters, the stellar evolution-driven 
abundance changes match well with what is expected from the literature. 
Surface carbon depletion, a result of slow circulation of material between 
the photosphere and the hydrogen-burning shell in bright red giants, begins 
near the ``bump'' in the RGB luminosity function and proceeds at a rate 
consistent with the observations of Martell et al. (2008c)\nocite{MSB08c} 
and the predictions of Denissenkov \& VandenBerG (2003)\nocite{DV03}. 

The data set presented here is particularly well suited to the study of deep 
mixing in the red giants in NGC 5466, since the data extends to quite faint 
magnitudes. This is made possible by the multiplexing ability of VIRUS-P as 
an integral field spectrograph. We also show that 
VIRUS-P can be used to get velocities for stars to better than 17 \kms
if good template matching is done.   The future VIRUS instrument on
the Hobby-Eberly Telescope will have comparable spectral resolution.
However, instead of a single spectrograph 
it will have more than 100 IFU
spectrographs, and the fiber size on the sky will shrink from 
$4.1\arcsec$ to $1.5\arcsec$.
This will make VIRUS a very capable instrument for observing larger, 
more crowded globular clusters.

\acknowledgments
Support for R.W. was provided by the National Science Foundation 
through AST-0649128, which funds the McDonald Observatory REU
program.

Support for S.L.M. and G.H.S. was provided by the NSF grant AST-0406988.

The authors recognize and acknowledge the very significant cultural role and reverence that the summit of Mauna Kea has always had within the indigenous Hawaiian community.  We are most fortunate to have the opportunity to conduct observations from this mountain.


\begin{deluxetable}{lrrrrrcc}
\tablewidth{0pt}
\tablecaption{Velocity Sample}
\tablehead{
\colhead{ID} &
\colhead{$\alpha_{2000}$} &
\colhead{$\delta_{2000}$} &
\colhead{$B$} &
\colhead{$V$} &
\colhead{$I$} &
\colhead{RV ($km s^{-1}$)} &
\colhead{$\sigma_{\mathrm{RV}}$ ($km s^{-1}$)} }
\startdata
\multicolumn{8}{c}{NGC 5466 members}\\
3209 & 14:05:23.45 & 28:30:01.8 & 16.84 & 16.51 & 16.03 &  86.5  &  2.6\\
1409 & 14:05:24.49 & 28:32:44.7 & 16.47 & 16.17 & 15.74 &  87.9  &  6.1\\
2176 & 14:05:25.76 & 28:31:44.6 & 17.01 & 16.60 & 16.12 & 107.9  & 11.3\\
2481 & 14:05:26.39 & 28:31:20.3 & 15.52 & 14.55 & 13.47 & 112.7  &  8.2\\
1765 & 14:05:27.62 & 28:32:17.7 & 16.75 & 16.61 & 16.52 & 100.1  &  2.7\\
2245 & 14:05:28.05 & 28:31:39.7 & 16.78 & 16.61 & 16.39 & 105.6  &  9.5\\
1578 & 14:05:29.62 & 28:32:33.0 & 16.75 & 16.68 & 16.56 & 103.9  &  4.6\\
1899 & 14:05:30.42 & 28:32:08.5 & 16.78 & 16.39 & 15.87 &  81.8  & 13.7\\
2281 & 14:05:31.92 & 28:31:37.8 & 17.16 & 16.40 & 15.54 & 103.7  &  3.7\\
1524 & 14:05:34.48 & 28:32:39.1 & 17.16 & 16.74 & 16.20 &  93.4  &  4.0\\
1235 & 14:05:36.93 & 28:33:05.7 & 16.92 & 16.11 & 15.20 & 110.8  &  9.4\\
2753 & 14:05:19.35 & 28:30:54.9 & 17.85 & 17.13 & 16.33 & 151.8  &  7.6\\
2840 & 14:05:24.90 & 28:30:48.0 & 18.01 & 17.31 & 16.47 & 97.1   & 10.6\\
2113 & 14:05:26.18 & 28:31:49.8 & 16.96 & 16.44 & 15.73 & 130.2  & 11.5\\
1483 & 14:05:30.45 & 28:32:40.8 & 16.90 & 16.49 & 15.90 & 145.3  &  6.2\\
3305 & 14:05:20.71 & 28:29:41.8 & 15.67 & 14.67 & 13.58 & 128.7  &  3.5\\
3127 & 14:05:22.82 & 28:30:13.8 & 14.88 & 13.56 & 12.21 & 122.8  &  6.7\\
3342 & 14:05:23.30 & 28:29:35.6 & 16.74 & 16.67 & 16.51 & 122.0  &  5.7\\
2891*& 14:05:23.54 & 28:30:41.7 & 15.60 & 14.57 & 13.48 & 111.2  &  4.8\\
1839 & 14:05:24.86 & 28:32:11.1 & 15.39 & 14.13 & 12.95 & 116.5  &  3.6\\
1461 & 14:05:24.91 & 28:32:40.5 & 17.12 & 16.37 & 15.43 & 123.0  &  9.5\\
1749 & 14:05:25.12 & 28:32:18.0 & 17.56 & 16.83 & 15.99 & 106.3  &  4.5\\
3401 & 14:05:25.60 & 28:29:22.9 & 18.40 & 17.81 & 16.97 & 113.3  &  2.0\\
3071 & 14:05:25.37 & 28:30:23.1 & 15.53 & 14.54 & 13.47 & 118.1  & 10.6\\
3033 & 14:05:25.86 & 28:30:27.4 & 16.40 & 15.84 & 15.15 & 119.2  & 11.2\\
2821 & 14:05:25.77 & 28:30:50.1 & 16.35 & 15.52 & 14.54 & 117.9  &  7.2\\
2919 & 14:05:26.27 & 28:30:40.4 & 17.07 & 16.33 & 15.43 & 117.1  &  4.1\\
2776 & 14:05:26.63 & 28:30:55.2 & 16.23 & 15.39 & 14.41 & 108.1  &  4.1\\
1351 & 14:05:27.14 & 28:32:51.2 & 17.20 & 16.42 & 15.53 & 123.4  &  8.1\\
1946 & 14:05:28.29 & 28:32:04.3 & 17.54 & 16.82 & 15.95 & 109.5  & 12.6\\
2440 & 14:05:30.17 & 28:31:25.0 & 16.40 & 15.56 & 14.58 & 121.8  &  0.5\\
2544 & 14:05:30.90 & 28:31:16.1 & 16.69 & 15.84 & 14.94 & 133.0  & 18.8\\
1497 & 14:05:31.03 & 28:32:39.9 & 15.92 & 14.99 & 13.94 & 112.4  &  5.5\\
2483 & 14:05:31.19 & 28:31:21.8 & 15.75 & 14.85 & 13.85 & 121.8  &  8.9\\
1972 & 14:05:31.38 & 28:32:02.8 & 15.69 & 14.68 & 13.61 & 104.5  &  2.6\\
1615 & 14:05:31.86 & 28:32:30.7 & 18.53 & 17.93 & 17.12 & 113.5  &  3.5\\
 826 & 14:05:32.65 & 28:33:51.6 & 17.40 & 16.69 & 15.85 & 122.2  & 16.1\\
1534 & 14:05:33.09 & 28:32:37.8 & 16.06 & 15.16 & 14.14 & 121.6  &  7.1\\
 904 & 14:05:33.09 & 28:33:41.7 & 15.23 & 14.06 & 12.83 & 123.6  &  7.8\\
 707 & 14:05:33.53 & 28:34:08.6 & 16.71 & 15.91 & 14.98 & 130.2  &  8.5\\
 727*& 14:05:34.44 & 28:34:06.2 & 15.50 & 14.46 & 13.33 & 136.9  &  6.7\\
1361 & 14:05:34.84 & 28:32:52.5 & 18.50 & 17.83 & 17.05 & 160.8  & 13.4\\
1075 & 14:05:34.96 & 28:33:20.6 & 17.85 & 17.14 & 16.29 & 133.8  & 17.1\\
1320*& 14:05:35.22 & 28:32:56.7 & 15.62 & 14.61 & 13.51 & 131.5  &  6.3\\
1024 & 14:05:35.55 & 28:33:27.4 & 18.20 & 17.57 & 16.74 & 129.3  &  5.1\\
1112 & 14:05:36.00 & 28:33:16.9 & 18.33 & 17.65 & 16.86 & 112.4  &  3.9\\
 690 & 14:05:39.51 & 28:34:14.4 & 17.92 & 17.22 & 16.42 & 105.8  &  0.6\\
2473 & 14:05:39.92 & 28:31:25.2 & 16.73 & 16.35 & 15.77 & 109.4  & 13.0\\
2635*& 14:05:44.53 & 28:31:13.5 & 15.56 & 14.62 & 13.55 & 103.0  &  5.6\\
2629 & 14:05:39.28 & 28:31:12.9 & 14.89 & 13.61 & 12.27 & 108.1  &  7.0\\
2642 & 14:05:42.19 & 28:31:12.1 & 16.76 & 16.71 & 16.63 & 111.0  & 13.0\\
3176*& 14:05:39.96 & 28:30:11.1 & 15.61 & 14.60 & 13.50 & 133.7  & 13.0\\
3306 & 14:05:41.08 & 28:29:48.2 & 15.90 & 14.97 & 13.92 & 131.7  &  6.0\\
837  & 14:05:39.30 & 28:33:52.3 & 17.13 & 16.36 & 15.46 &  121.2  &  6.7\\
1398 & 14:05:29.10 & 28:32:47.3 & 15.32 & 14.48 & 13.48 &  123.2  &  7.4\\
1344 & 14:05:26.71 & 28:32:51.6 & 18.66 & 18.20 & 17.34 &  103.1  & 41.7\\
1106 & 14:05:32.58 & 28:33:16.2 & 18.89 & 18.40 & 17.63 &   75.8  & 17.1\\
3387 & 14:05:20.09 & 28:29:24.1 & 19.21 & 18.66 & 17.88 &  132.8  &  6.9\\
2725 & 14:05:22.37 & 28:30:58.4 & 18.82 & 18.23 & 17.41 &  123.8  & 29.8\\
2957 & 14:05:22.72 & 28:30:33.9 & 18.09 & 17.64 & 16.82 &   94.9  &  9.4\\
3192 & 14:05:22.77 & 28:30:03.6 & 19.46 & 19.29 & 18.78 &  152.9  & 22.5\\
3078 & 14:05:23.48 & 28:30:21.4 & 19.25 & 18.67 & 17.90 &  144.6  & 14.4\\
946  & 14:05:33.66 & 28:33:36.3 & 19.03 & 18.46 & 17.71 &  108.8  & 15.2\\
968  & 14:05:35.90 & 28:33:33.7 & 19.49 & 18.88 & 18.16 &  113.8  &  5.1\\
888  & 14:05:36.45 & 28:33:45.2 & 18.64 & 18.08 & 17.28 &  115.3  &  8.1\\
3113 & 14:05:26.33 & 28:30:17.1 & 18.96 & 18.35 & 17.58 &  122.3  & 16.3\\
\multicolumn{8}{c}{Sample non-members}\\
1980 & 14:05:31.85 & 28:32:02.6 & 17.95 & 17.21 & 16.47 &  68.9  &  6.3\\
\enddata
\label{t1}
\end{deluxetable}

\clearpage
\begin{deluxetable}{lrrrrrr}
\tablecaption{Results}
\tablehead{
\colhead{ID} &
\colhead{$S(3839)$} &
\colhead{$\sigma_{S}$} &
\colhead{$S_{2}(CH)$ } &
\colhead{$\sigma_{S2}$} &
\colhead{ [C/Fe] } &
\colhead{$\sigma_{\mathrm{[C/Fe]}}$} }
\startdata
\multicolumn{7}{c}{Main sample}\\
1839 &  0.460 & 0.027 & 0.793 & 0.038 &  0.02 & 0.18    \\
1398 & -0.185 & 0.034 & 0.603 & 0.042 & -0.68 & 0.18     \\
2483 & -0.199 & 0.026 & 0.609 & 0.039 & -0.81 & 0.14     \\
2629 & -0.093 & 0.055 & 0.510 & 0.046 & -1.14 & 0.26     \\
3176 & -0.028 & 0.033 & 0.622 & 0.041 & -0.69 & 0.13     \\
3306 & -0.160 & 0.031 & 0.663 & 0.040 & -0.60 & 0.43     \\
2635 & -0.144 & 0.031 & 0.596 & 0.040 & -0.87 & 0.17     \\
3127 & -0.040 & 0.024 & 0.522 & 0.037 & -0.91 & 0.13     \\
 904 & -0.087 & 0.023 & 0.591 & 0.038 & -0.80 & 0.18    \\
 727 & -0.135 & 0.025 & 0.648 & 0.038 & -0.49 & 0.14    \\
3071 & -0.166 & 0.037 & 0.587 & 0.042 & -0.86 & 0.13     \\
2891 & -0.110 & 0.025 & 0.636 & 0.038 & -0.58 & 0.11     \\
1320 & -0.139 & 0.024 & 0.630 & 0.038 & -0.58 & 0.11     \\
3305 & -0.144 & 0.026 & 0.658 & 0.038 & -0.48 & 0.12     \\
1497 & -0.102 & 0.026 & 0.645 & 0.038 & -0.52 & 0.11     \\
1534 & -0.166 & 0.024 & 0.648 & 0.038 & -0.47 & 0.11        \\
2776 & -0.145 & 0.026 & 0.664 & 0.039 & -0.39 & 0.11        \\
2821 & -0.209 & 0.029 & 0.688 & 0.040 & -0.25 & 0.12        \\
2440 & -0.116 & 0.029 & 0.638 & 0.040 & -0.52 & 0.11        \\
2544 & -0.218 & 0.028 & 0.656 & 0.040 & -0.42 & 0.13        \\
 707 & -0.206 & 0.026 & 0.709 & 0.039 & -0.08 & 0.12       \\   
1749 & -0.168 & 0.039 & 0.673 & 0.045 & -0.11 & 0.13     \\
 826 & -0.138 & 0.034 & 0.685 & 0.042 & -0.05 & 0.10    \\
1946 & -0.185 & 0.037 & 0.701 & 0.044 & -0.05 & 0.11     \\
3401 & -0.203 & 0.032 & 0.685 & 0.041 & -0.01 & 0.14     \\
 690 & -0.265 & 0.054 & 0.736 & 0.054 &  0.16 & 0.15    \\
2753 & -0.239 & 0.044 & 0.709 & 0.048 & -0.05 & 0.21     \\
 837 & -0.214 & 0.031 & 0.710 & 0.041 &  0.02 & 0.13    \\
2957 & -0.238 & 0.091 & 0.660 & 0.079 &  0.12 & 0.12     \\
\multicolumn{7}{c}{Sample with potential blends}\\
2919 & -0.209 & 0.042 & 0.726 & 0.046 &  0.01 & 0.12     \\
1351 & -0.188 & 0.031 & 0.712 & 0.041 & -0.05 & 0.14     \\
1075 & -0.220 & 0.036 & 0.704 & 0.044 &  0.05 & 0.11     \\
1361 & -0.184 & 0.053 & 0.663 & 0.053 & -0.03 & 0.13     \\
1615 & -0.215 & 0.036 & 0.685 & 0.043 &  0.06 & 0.10     \\
1024 & -0.191 & 0.038 & 0.659 & 0.045 & -0.12 & 0.15     \\
2481 & -0.174 & 0.039 & 0.560 & 0.043 & -0.96 & 0.11     \\
1972 & -0.088 & 0.066 & 0.631 & 0.055 & -0.56 & 0.11     \\
\enddata
\label{t2}
\end{deluxetable}

\clearpage


\end{document}